\documentclass[compsoc,conference,a4paper,10pt,times]{IEEEtran}
\IEEEoverridecommandlockouts
\usepackage[hyphens,spaces,obeyspaces]{url}
\usepackage{cite}
\usepackage{amsmath,amssymb,amsfonts}
\usepackage{algorithmic}
\usepackage{graphicx}
\usepackage{textcomp}
\usepackage{bmpsize}
\usepackage{xcolor}
\usepackage{lipsum}
\usepackage{multirow}
\usepackage{pifont}
\usepackage[colorlinks=true,urlcolor=black]{hyperref}
\def\BibTeX{{\rm B\kern-.05em{\sc i\kern-.025em b}\kern-.08em
    T\kern-.1667em\lower.7ex\hbox{E}\kern-.125emX}}
    
\usepackage{xparse}
\NewDocumentCommand\mm{g}{%
  \IfNoValueF{#1}{{\color{orange} \textbf{(MM: #1)}}}%
  \IfNoValueT{#1}{{\color{orange} \textbf{(MM)}}}%
}

\NewDocumentCommand\mk{g}{%
  \IfNoValueF{#1}{{\color{red} \textbf{(MK: #1)}}}%
  \IfNoValueT{#1}{{\color{red} \textbf{(MK)}}}%
}

\NewDocumentCommand\kb{g}{%
  \IfNoValueF{#1}{{\color{ACMDarkBlue} \textbf{(KB: #1)}}}%
  \IfNoValueT{#1}{{\color{ACMDarkBlue} \textbf{(KB)}}}%
}

\NewDocumentCommand\bk{g}{%
  \IfNoValueF{#1}{{\color{green} \textbf{(BK: #1)}}}%
  \IfNoValueT{#1}{{\color{green} \textbf{(BK)}}}%
}
    
\newcommand{\subhead}[1]{\vspace {1pt}\noindent{\textbf{#1.}}}

\newcommand{\BigMAC}[0]{\textsc{BigMAC}}
\newcommand{\ourtool}[0]{\textsc{Sausage}}
\newcommand{\datasetsize}[0]{200}
    
\begin{document}







\title{SAUSAGE: \\Security Analysis of Unix domain Socket usAGE in Android}

\makeatletter
\newcommand{\linebreakand}{%
  \end{@IEEEauthorhalign}
  \hfill\mbox{}\par
  \mbox{}\hfill\begin{@IEEEauthorhalign}
}
\makeatother

\author{\IEEEauthorblockN{Mounir Elgharabawy}
    \IEEEauthorblockA{
    \textit{Concordia University} \\
    Montreal, Quebec, Canada \\
    m\_elghar@encs.concordia.ca}
    \and
    \IEEEauthorblockN{Blas Kojusner}
    \IEEEauthorblockA{
    \textit{University of Florida} \\
    Gainesville, Florida, USA \\
    bkojusner@ufl.edu}
    \and
    \IEEEauthorblockN{Mohammad Mannan}
    \IEEEauthorblockA{
    \textit{Concordia University} \\
    Montreal, Quebec, Canada \\
    m.mannan@concordia.ca}
    \linebreakand
    \IEEEauthorblockN{Kevin R. B. Butler}
    \IEEEauthorblockA{
    \textit{University of Florida} \\
    Gainesville, Florida, USA \\
    butler@ufl.edu}
    \and
    \IEEEauthorblockN{Byron Williams}
    \IEEEauthorblockA{
    \textit{University of Florida} \\
    Gainesville, Florida, USA \\
    byron@cise.ufl.edu}
    \and
    \IEEEauthorblockN{Amr Youssef}
    \IEEEauthorblockA{
    \textit{Concordia University} \\
    Montreal, Quebec, Canada \\
    amr.youssef@concordia.ca}
}

\maketitle

\begin{abstract}
The Android operating system is currently the most popular mobile operating system in the world. Android is based on Linux and therefore inherits its features including its Inter-Process Communication (IPC) mechanisms. These mechanisms are used by processes to communicate with one another and are extensively used in Android. While Android-specific IPC mechanisms have been studied extensively, Unix domain sockets have not been examined comprehensively, despite playing a crucial role in the IPC of highly privileged system daemons. In this paper, we propose \ourtool, an efficient novel static analysis framework to study the security properties of these sockets. \ourtool~considers access control policies implemented in the Android security model, as well as authentication checks implemented by the daemon binaries. It is a fully static analysis framework, specifically designed to analyze Unix domain socket usage in Android system daemons, at scale. We use this framework to analyze \datasetsize~Android images across eight popular smartphone vendors spanning Android versions 7-9. As a result, we uncover multiple access control misconfigurations and insecure authentication checks. Our notable findings include a permission bypass in highly privileged Qualcomm system daemons and an unprotected socket that allows an untrusted app to set the scheduling priority of other processes running on the system, despite the implementation of mandatory SELinux policies. Ultimately, the results of our analysis are worrisome; all vendors except the Android Open Source Project (AOSP) have access control issues, allowing an untrusted app to communicate to highly privileged daemons through Unix domain sockets introduced by hardware manufacturer or vendor customization.
\end{abstract}


\section{Introduction}
One of the fundamental features any modern operating system provides is Inter-Process Communication (IPC), used extensively by applications to implement inter-functionality between their components. The widely-used Android OS provides a variety of its own IPC mechanisms (e.g., Binder, Intents, Messenger), while also inheriting the traditional IPC mechanisms available in a Linux environment, in particular, Unix domain sockets. As with any IPC mechanism, privileged processes communicating over these sockets are potentially vulnerable to confused deputy attacks, if they are inadequately protected by the access control policy. In that case, a malicious unprivileged process can control a privileged process to overwrite critical files~\cite{weaksauce}, execute shell commands~\cite{CVE-2013-4777}, or gather screenshots and sensitive system logs~\cite{aee_vuln}, among other things.


Multiple vulnerabilities and exploits are due to the misuse of Unix domain sockets. For example, CVE-2011-3918~\cite{CVE-2011-3918} describes an unprotected socket to the Zygote process, which can be leveraged to perform a denial of service attack on a device running AOSP Android 4.0.3. Other examples include the ``HTC WeakSauce'' exploit, which uses a socket connection to the privileged \textit{dmagent} system daemon to achieve privilege escalation to root~\cite{weaksauce}, and  CVE-2013-4777 and CVE-2013-5933~\cite{CVE-2013-4777, CVE-2013-5933}, privilege escalation vulnerabilities affecting Motorola devices, due to an unprotected socket to the \textit{init} process. More recently, an information disclosure vulnerability was discovered on Huawei phones, allowing attackers to gather screenshots and kernel and system logs~\cite{aee_vuln}. The vulnerability is exploited by sending commands via an exposed socket to a vendor-customized version of the \textit{debuggerd} daemon. These vulnerabilities demonstrate that unprotected sockets can degrade the security of the system whether they originate in stock (i.e., AOSP), or vendor-customized Android. 

However, previous research has primarily focused on Android-specific IPCs, such as Binder~\cite{binderfuzz1, binderfuzz2, fans, inputvalidation, chizpurfle} and Intents~\cite{pendingintents, intentfuzzer}, with comparatively little evaluation of traditional Linux IPCs such as Unix domain sockets. While Shao et al.~\cite{uds} examined the misuse of Unix domain sockets in Android, finding that the inadequate protection of these sockets is a common pitfall in Android, their approach fell short for system daemons, which are arguably much more valuable targets for exploitation. This shortcoming is due to the use of dynamic analysis to avoid challenges such as reasoning about the complex interaction of Android access control layers, extracting firmware images from different vendors with different formats, and statically analyzing system daemon ARM binaries accurately. As a consequence, their analysis requires access to a running, rooted Android device, and thus only covers two vendors across three Android versions. Also,  the approach of Shao et al.\ makes a cross-vendor analysis infeasible at scale, despite being essential to uncover the misuse of sockets introduced by vendor customization. Furthermore, it fails to uncover inactive sockets, which can be created in response to an event or configuration change.

To address this gap, we propose \ourtool, a static analysis framework to identify valid socket connections that untrusted apps can establish to system daemons on an Android device, without the need of a running device, enabling large-scale analysis. Given an Android firmware image, our framework analyzes access control policies and performs static binary analysis on daemon binaries to discover socket addresses that an untrusted app can connect to and any authentication checks implemented in the binary. We overcome the challenge of reasoning about Android access control policies by using a version of the \BigMAC~\cite{bigmac} SELinux policy analysis tool; we extend its functionality to enable socket creation within the init boot simulation step. Since all Android IPC is governed by MAC (and sometimes by DAC), we believe that our approach may serve as a good base for future work examining the security of Android IPC mechanisms (especially statically). We also implement our own binary analysis component. The \ourtool~framework extracts the system's SELinux policy, system daemon binaries and init RC files from an Android firmware image. It analyzes the SELinux policy to determine which system daemons an untrusted app can communicate with. By using inter-procedural data-flow analysis, it then detects socket addresses, their access control credentials, and any authentication checks in the system daemon binaries with high accuracy. 

We used our framework to analyze \datasetsize~Android firmware images, spanning eight different vendors and Android versions 7-9. \ourtool{} fully analyzes a firmware image in around 14 minutes, making it scalable as a cross-vendor Android firmware analysis tool, without requiring vendor-specific devices. The results of our analysis are worrisome; all vendors except AOSP have access control issues that allow an untrusted app to communicate to highly privileged daemons. These include HTC \textit{dmagent} that has been previously exploited in ``HTC WeakSauce,'' and Samsung's Professional Audio service, which allows any app to set its process scheduling priority.  We also identify insecure authentication practices used by these daemons, such as checks based on an app's process name, which can be trivially spoofed.  Additionally, we demonstrate that our approach can uncover Unix domain sockets that would have been missed by the dynamic analysis approach used by Shao et al.~\cite{uds}



%
\subhead{Contributions}

\begin{enumerate}

\item We propose an access control-aware, fully static framework to analyze Unix domain socket usage in Android system daemons, compatible with Android versions 7.0 and above. Using this framework, we conduct an analysis of \datasetsize{} Android factory images spanning versions 7.0-9.0 across eight different vendors, including prominent players such as Samsung and Xiaomi, which account for over a third of the mobile vendor market share worldwide~\cite{MobileVe46:online}, and others such as Asus, Motorola, HTC, and AOSP, to find system daemon sockets accessible to untrusted apps.\footnote{We use the term \textit{untrusted app} to reference third-party applications a user can install.} We compare our results to the ground truth from three running devices and find that our framework achieves 100\% accuracy in detecting socket addresses and their MAC and DAC credentials.
\item We use a novel methodology based on a version of \BigMAC{} (that we modified at the init boot simulation step) to reason about Android's complex interaction of access control layers through static analysis. We will publish the source of the binary analysis module to the community, as well as the modified version of \BigMAC{} we use as part of our framework.
\item We find multiple instances of unprotected Unix domain sockets to root processes that could lead to exploits such as HTC WeakSauce~\cite{weaksauce}. Our approach detects Unix domain sockets that are created under certain conditions and would not have been detected through past approaches relying on dynamic analysis alone.

\end{enumerate}

\subhead{Notable Findings}

\begin{enumerate}
    \item We found two highly privileged Qualcomm system daemons, \textit{cnd} and \textit{dpmd}, where a faulty authentication mechanism is used to authenticate the peer, relying only on its process name. However, the process name can be easily spoofed in both cases. Through static analysis, we infer that this allows clients to get/set network settings such as WiFi AP, WiFi P2P, and Default Network settings, by sending the appropriate command over the cnd socket. 
    
    \item In 25 Samsung Android 7.0-7.1 images, we found that the daemon apaservice listens over a socket that can be used to request changing the scheduling priority for any process to any priority. This can result in DoS of the Samsung audio subsystem. Additionally, the daemon is vulnerable to buffer overflow, allowing an untrusted app to execute code in the daemon's security context. This vulnerability was assigned CVE-2021-25461.
    
    \item There are 
    multiple instances of overly permissive SELinux policies in seven of the eight vendors we analyzed. These policies allow socket communication between untrusted apps and highly privileged system daemons, weakening the system's overall security posture. Examples of these daemons include \textit{dumpstate} and \textit{rild} in HTC, and \textit{cnd} in most vendors.
    
    \item We discovered 
    multiple instances of vendor customization of AOSP binaries that expose additional unprotected sockets. One example is HTC \textit{rild} where two custom sockets were added that are configured to be accessible to an untrusted app.
    
\end{enumerate}

We have contacted Samsung and Qualcomm as part of our responsible disclosure process with details of the vulnerabilities and proof-of-concepts. We detail this process in Section~\ref{sec:responsible_disclosure}.

\subhead{Outline} The remainder of the paper is structured as follows: Section 2 provides the necessary background on how Unix domain sockets work and their role in the Android security model. Section 3 details key related work. Section 4 develops the design of \ourtool~and Section 5 its implementation. In Section 6 we evaluate \ourtool~against multiple Android firmware images and demonstrate the scalability and impact it can have when discovering accessible sockets. In Section 7 we present case studies on the results obtained with \ourtool.  We discuss our limitations in Section 8, and finally conclude in Section 9.


\section{Background}

In this section, we provide background about how Unix domain sockets work and where they fit in the Android security model. 

\subsection{Android Security Model}

The Android security model implements various layers. Apps run in sandboxes defined by the creation of a unique Linux UID for each application at install time. Processes can only communicate with other applications via enforced mandatory access control (MAC). This is a feature implemented in Security-Enhanced Linux (SELinux) through the modified SEAndroid framework. Interested readers can find a comprehensive discussion of the Android security model in~\cite{android_security_model}. In this section, we explain the key concepts behind it that are relevant to our work.

\subhead{Discretionary Access Control on Android (DAC)} DAC is an access control model that is used by Linux. It is implemented in Android by using a fixed set of user and group IDs for system-related purposes and limiting a range of user IDs for dynamically installed applications. Android limits the number of processes that can run as root, therefore a highly-privileged process would typically run under the system UID, or another UID specific to execute the role intended for the process. This avoids granting more permissions than necessary to a process, which can inherently avoid security issues. Untrusted apps are assigned a unique UID from a specified range of IDs that are available. This prevents third-party applications from having more access than necessary on any other files that are not included in their installation.

\subhead{SELinux} Security Enhanced Linux (SELinux) provides a framework for enforcing Mandatory Access Control (MAC) on Linux. 
It was introduced into the Android platform  in 2013 through the SEAndroid framework~\cite{smalley2001implementing}. SELinux contains a set of rules that are based on file labels which contain information such as user, role, type, and level. These rules determine what types and actions a process has access to and are structured to group items together based on their accessibility. In Android, SELinux is not only restricted to access control for files, but it also manages access control for IPC mechanisms, such as Binder and Unix domain sockets. Thus, for processes to communicate with one another, the communication must be explicitly allowed by the SELinux policy.

SELinux policies are developed by vendors by combining the core AOSP policy with device-specific customization. Policies compare rules that guide the SELinux security engine, including types for file objects and domains for processes. The SELinux policy uses roles to limit the domains that can be accessed and has user identities to specify the roles that users can have. New rules can be added into the policy which is then preprocessed and built into the policy.conf file.


\subhead{Supplementary Groups} Adding a supplementary group ID to an application will grant it all the privileges of the specified group. The groups for an application are assigned within the manifest file. An example of some supplementary groups would be the Bluetooth group or the Internet group. The permissions of a supplementary group can be enforced at the kernel level or at the Android Framework level depending on the functionality granted to the group~\cite{ayed2015literature}.

\subhead{Middleware Permissions} The Android middleware layer contains a reference monitor that mediates inter-component communication~\cite{understandingAndroidSec}. Middleware permissions grant apps access to resources and services that are provided by the Android operating system rather than the Linux kernel.

\subsection{Unix Domain Sockets}

A Unix domain socket is a communications endpoint for exchanging data between processes on the same host operating system. It can also be referred to as an inter-process communication socket. 
The main difference between Unix domain sockets and Internet sockets is that a Unix domain socket is an IPC where all communication occurs strictly within the operating system kernel. Internet sockets use an underlying network protocol for communication. Unix domain sockets also have what is called a namespace, or a unique identifier to an object of a certain kind, that is used to label the socket types. There are three types of Unix domain socket namespaces in Android, as can be seen in Table \ref{table:namespaces}.

\begin{table}[t]
\caption{Security enforcement corresponding to Android Unix domain socket namespaces 
}
\centering
\label{table:namespaces}
\begin{tabular}{|l|c|c|}
\hline
\multicolumn{1}{|c|}{\multirow{2}{*}{\textbf{Namespace}}} & \multicolumn{2}{c|}{\textbf{Security Enforcement}} \\ \cline{2-3} 
\multicolumn{1}{|c|}{}                                    & \textbf{SELinux}    & \textbf{File Permissions}    \\ \hline
RESERVED   & \ding{51} & \ding{51} \\ 
FILESYSTEM & \ding{51} & \ding{51} \\ 
ABSTRACT   & \ding{51} & \ding{55}  \\ \hline
\end{tabular}
\end{table}

\subhead{FILESYSTEM} Sockets with this namespace are associated with a file on the file system and are created by a process that needs them. Once a socket file is created it will be protected by the discretionary access control system, or the DAC, as well as the mandatory access control, or the MAC. Only processes with the proper read and write permissions can communicate with these socket files.

\subhead{RESERVED} This namespace is introduced in Android and falls under the FILESYSTEM namespace and thus inherits its access control properties. The socket files are created by init and are located under \texttt{/dev/socket}. The name indicates that these socket files are reserved for system use. The socket file descriptors are made available to their owner service daemon through an environment variable named \texttt{ANDROID\_SOCKET\_<addr>} where \texttt{<addr>} is the address of the socket in the RESERVED namespace.

\subhead{ABSTRACT} These sockets allow a program to bind a Unix domain socket to a name without the name being created in the filesystem. The socket's name begins with a null byte which removes the need to create a filesystem path name for the socket. 
 
There are three prerequisites for a process to be allowed to establish a connection to a FILESYSTEM socket. First, the connecting process must be allowed to communicate to the server process through the Unix domain socket IPC by SELinux. Second, the connecting process must be allowed to write to the socket file, based on its file context in SELinux. Third, the connecting process must have the appropriate UID or GID to write to the socket file, depending on the socket file's DAC file permissions. On the other hand, only the first prerequisite is needed in the case of ABSTRACT sockets since file-based access control policies are not applicable to them. As a result, ABSTRACT sockets are the least secure of the three namespaces. Furthermore, Unix domain sockets can only be bound by one process. Filesystem MAC and DAC can restrict the creation of sockets under certain directories to a set of processes, preventing untrusted apps from binding sockets used by system daemons. This does not apply to ABSTRACT sockets, however, allowing a malicious app to DoS the system daemon by occupying an ABSTRACT address if the app manages to bind the socket address before the daemon. Fortunately, daemons started by init are always started before apps, and in most cases, their sockets are bound on initialization and stay bound for the daemon's entire lifecycle.





\section{Related Work}

Android IPC security has long been the focus of a large body of research. Most of this research, however, centered on either the Binder IPC interface~\cite{binderfuzz1, binderfuzz2, fans, inputvalidation, chizpurfle}, or on Android Intents~\cite{pendingintents, intentfuzzer}. Furthermore, the majority of these analyses are concerned with Android application security rather than Android framework security. Iannillo et al.~\cite{chizpurfle} designed \textit{Chizpurfle}, a grey-box fuzzer for system services that discovers vendor-specific system service methods exposed through Binder IPC and runs a fuzzing campaign on the identified methods. However, it heavily relies on analyzing Java reflection by design, so it is not compatible with native system services. Liu et al.~\cite{fans} overcame this limitation with, FANS, which is capable of finding vulnerabilities in native system services, but is limited to examining the Binder IPC mechanism. While, these works clearly demonstrate their effectiveness finding vulnerabilities in system services, there is no clearly defined threat model that accounts for the multi-layered Android security model. Thus, some of the reported vulnerabilities from these works may require chaining with other privilege escalation exploits to interact with privileged, but vulnerable, Binder interfaces. Additionally, the chaining may be prevented by commonly deployed access control policies.

Shao et al.~\cite{uds} conducted the first study of Unix domain socket usage by both Android apps and system daemons. To perform their analysis, they developed SInspector, which identifies Unix domain socket addresses and detects authentication checks. SInspector exclusively utilizes static techniques for apps, allowing for large scale analysis of apps. However, for system daemons, the tool needs to be run on a live, rooted Android system. The reasons for this limitation include the difficulty of unpacking Android factory images from different vendors and the complexity of security enforcement for sockets, which involves the interplay of SEAndroid and DAC file permissions. However, with the advent of new open-source tools such as the Android image unpacking library~\cite{at_commands} and \BigMAC{}~\cite{bigmac}, it is now feasible to tackle these challenges and develop a fully static large-scale cross-vendor analysis framework for Unix domain socket usage in system daemons. We use these tools to overcome the challenges faced by SInspector that caused it to resort to dynamic analysis for system daemons.

A separate growing body of work examines Android OS security from an access control perspective~\cite{polyscope, bigmac, easeandroid, sepal, trustbutverify}. Android access control, especially SEAndroid, plays an essential role in securing IPC. All Android IPC mechanisms are protected by the SEAndroid policy and some are protected by DAC. Lee et al. proposed PolyScope~\cite{polyscope}, a tool to vet Android filesystem access control policies. They define three possible patterns of integrity violations in access control policies and rely on AOSP documentation as well as the integrity walls method~\cite{integrity_walls} to categorize process into different integrity levels. However, PolyScope requires a rooted phone, precluding the possibility of using it in a static analysis framework. Hernandez et al.~\cite{bigmac} proposed \BigMAC{} that combines DAC and MAC to construct an attack graph representing allowed data-flows between subjects and objects in a running system. \BigMAC{} succeeds to recover the running system's security state purely though static analysis with high accuracy. This makes it an ideal candidate as a base on which we can bootstrap more in-depth analysis. Although \BigMAC{} serves to abstract away the complexity of the Android security model, which was recently detailed by \cite{android_security_model}, it cannot detect DAC checks that occur dynamically in a running process.





\section{Design}

\begin{figure*}[t]
    \centering
    \includegraphics[width=\textwidth]{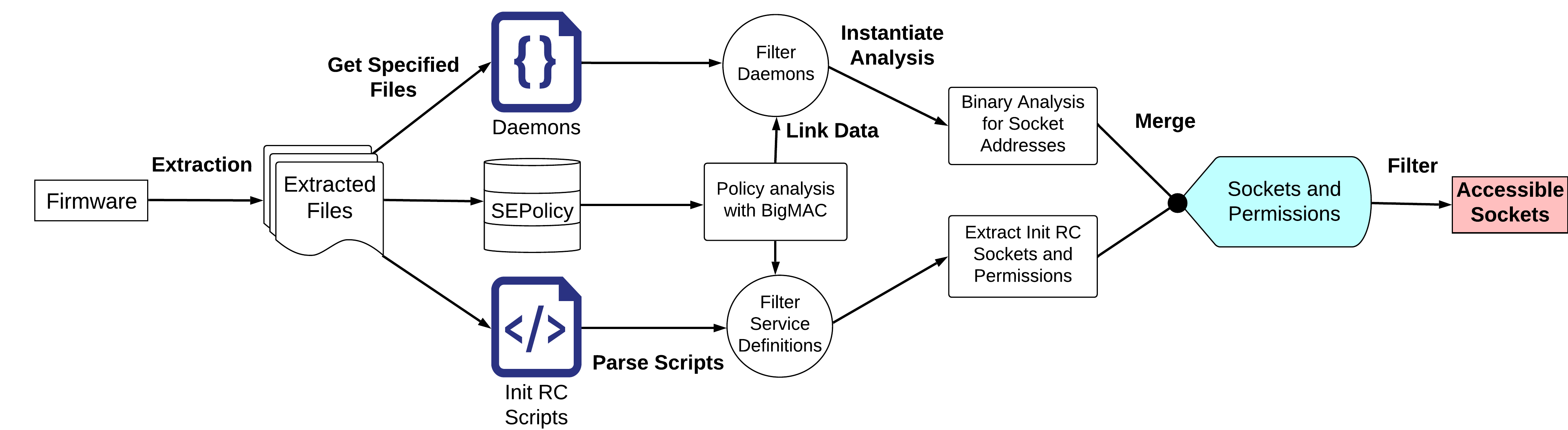}
    \caption{\ourtool~Framework Architecture. Following the unpacking of an image, the SELinux policy is analyzed to find the processes an untrusted app can communicate with via Unix domain sockets. The processes' binaries and their RC service definitions are retrieved from the extracted filesystem and analyzed independently to find the socket addresses that the process uses, their access control configurations, and any authentication checks in the binary code. The list of socket addresses is filtered down to the ones an untrusted app can access based on their access control permissions.}
    \label{fig:design}
\end{figure*}

An overview of the architecture of the tool can be seen in Figure \ref{fig:design}, we describe these steps in further detail below. The tool begins with the firmware image extraction. The filesystem is unpacked and extracted to acquire the SELinux policy, daemon binaries, and the init RC scripts. The SELinux policy is analyzed using a modified version of \BigMAC~\cite{bigmac} to query the processes an untrusted app can communicate with via sockets. The results from the analysis are then compiled into a list that is separately filtered with the daemon binaries and the relevant service definitions we have been able to extract. Once filtered with the service binaries, the tool conducts binary analysis to verify the socket address and find any security checks that may be in the binaries. Running in parallel is the Init RC Service Definition Analysis which will extract socket addresses and file permissions from the init RC files. The output of the binary analysis is combined with the output of the Init RC Service Definition Analysis and compiled into a list that can be used to filter for accessible sockets.

\subsection{Threat Model}
Unix domain sockets can only be accessed from processes that have proper permissions when checked by the MAC and DAC. Our threat model specifically focuses on untrusted apps, labeled \texttt{untrusted\_app} in SELinux. In our threat model, an app is allowed to obtain middleware permissions that can be granted to any \texttt{untrusted\_app}. Middleware permissions are not part of the MAC or DAC, so they are not directly considered when a socket is being accessed. However, the DAC supplementary groups assigned to an untrusted app depend on the permissions it has. Thus, we assign the following supplementary group to an untrusted app in our threat model:

The \texttt{android.permission.INTERNET} permission, which corresponds to the \texttt{inet} group, allows the \texttt{untrusted\_app} to perform network operations, such as opening network sockets. The \texttt{android.permission.BLUETOOTH\_ADMIN} permission, which corresponds to \texttt{net\_bt\_admin}, allows applications to discover and pair Bluetooth devices. Similarly, the Bluetooth permission labeled  \texttt{android.permission.BLUETOOTH} permission, corresponding to \texttt{net\_bt}, allows applications to connect to paired Bluetooth devices. The final permission is the external storage permission, \texttt{android.permission.} \texttt{MANAGE\_EXTERNAL\_STORAGE}, which belongs to the \texttt{external\_storage} group, allows an application a broad access to external storage in Scoped storage, a feature in Android allowing an application to only have access to their application directory on external storage plus any media created by the app~\cite{Manifest42:online}.

\subsection{Image Extraction}

The initial step of our tool is the firmware image extraction from a repository of 
firmware images that we have collected. A majority of the images have either been downloaded directly from vendor specific websites, such as the AOSP firmware images website~\cite{AOSPDownload:online}, or from third-party websites such as  firmwarepanda~\cite{FirmwarePanda:online}. The image packing format varies by vendor, e.g., HTC required an RUU Decrypt tool to account for the \texttt{Rom Update Utility}
format, while Samsung required \texttt{LZ4} decompression. Thus, we rely on existing unified tools that support every Android vendor and version to avoid developing our own from scratch. We use a modified version of the ATextract tool~\cite{at_commands}
to extract the files needed to analyze the SELinux policies and native daemons from the firmware images. For Android versions higher than 8.0, and for vendors that are not supported by the modified version of the ATextract tool, we use the newly released unpacking tool developed by Possemato et al.~\cite{trustbutverify}. The tool was modified to restructure the extracted image in the right format to be interpreted by our analysis pipeline. Our analysis pipeline collects DAC/MAC/CAP metadata, init RC files, daemons, shared libraries and SELinux policy files to be used by \BigMAC{} and our binary analysis component.



\subsection{SELinux Policy Analysis}

Following extraction, our framework reasons about the system's access control policies in order to determine which processes an untrusted app can connect to through Unix domain sockets. The Android security model is based on the complex interaction of multiple security layers, including SEAndroid policies, Linux filesystem permissions and Linux capabilities. Thus, we use a version of \BigMAC{} that we modified, to extract and recreate the security state of the running system. \BigMAC{} is a fine-grained SELinux policy static analysis tool~\cite{bigmac}. It first goes through the filesystem and extracts files' DAC file permissions, SELinux labels and Linux capabilities. Then, it parses the system's init scripts and simulates commands that affect the filesystem (e.g., \texttt{mkdir}, \texttt{chmod}), as well as \texttt{service} commands which execute service binaries. Performing boot emulation is required to create files in the \texttt{/sys}, \texttt{/dev} and \texttt{/data} directories, which would not be present in a static firmware image. We extended \BigMAC{}'s init boot simulation step to parse the \texttt{socket} option of \texttt{service} commands in order to create the socket files in the /dev/socket/ directory. For a full discussion of \BigMAC{}, we refer readers to the original paper by Hernandez et al.~\cite{bigmac}.

Once the modified version of \BigMAC{} finishes analyzing an image and generating an attack graph, it provides a query engine that can be used to find all the objects an untrusted app can write to. We can filter the resultant list of objects to only include IPC objects of the ``socket'' type. Since each IPC object holds a reference to its owner process, we extract the file path of the process's binary executable. The next steps of the analysis are then performed on these binaries.

\subsection{Socket Address Extraction}

Once we have the set of binary files for processes that an untrusted app can communicate with through Unix domain sockets, we can start to extract the socket addresses that these processes are listening over. We employ two methods for each one of these service binaries: init RC parsing and static binary analysis. These two methods are complementary to each other; parsing init RC files guarantees all RESERVED socket addresses will be recovered, and static binary analysis will recover all three types of socket addresses that the binary might be listening over.

\subsubsection{Init RC Service Definitions}
For each one of these binaries, there exists one or more service definitions in the Android system's init RC files. These service definitions can have options that configure how and when \texttt{init} runs these files. One of these options, \texttt{socket},\footnote{The \texttt{socket} option follows the syntax: \texttt{socket <name> <type> <perm> [ <user> [ <group> [ <seclabel> ] ] ]} where \texttt{<name>} is the address of the socket, \texttt{<perm>}, \texttt{<user>} and \texttt{<group>} are its credentials on the filesystem, and \texttt{<seclabel>} is its SELinux label.} creates a socket file for the service in the RESERVED namespace, creates a file descriptor for this socket and binds it to the created socket file, and saves the socket's file descriptor as an environment variable\footnote{This environment variable's name is formatted as \texttt{ANDROID\_SOCKET\_<address>} where \texttt{<address>} is the address of the socket in the reserved namespace.} for later retrieval by the service process. Therefore, it is straightforward to retrieve all RESERVED socket addresses from service definitions, by finding and parsing the \texttt{socket} options. However, this method does not capture socket addresses in other namespaces.

\subsubsection{Static Binary Analysis}
In the binary analysis module, we first construct the Control Flow Graph (CFG) of the binary and all externally linked objects, and identify all defined functions. We then perform an inter-procedural dataflow analysis starting at the entry point of every function that calls the \texttt{bind} system call in the binary. At the \texttt{bind} callsite, we extract the value of the address argument. The address is checked to determine whether or not it is a Unix domain socket address. If it is, we detect the namespace that the address belongs to by checking the first character of that address. If it is a null byte, then it belongs to the ABSTRACT namespace and no further analysis is needed. If the address starts with a directory separator ('\texttt{/}'), then it belongs to the FILESYSTEM namespace, and we attempt to determine the permissions the socket file is created with. This is done by detecting all preceding \texttt{umask}, \texttt{seteuid} and \texttt{setegid} system calls in the binary and extracting their arguments. The same process is carried out for subsequent \texttt{chmod}, \texttt{fchmod}, \texttt{chown}, and \texttt{fchown} calls. Additionally, the static binary analysis module detects RESERVED socket addresses by performing the same type of dataflow analysis for functions that call \texttt{getenv}. If the requested environment variable name starts with the ``\texttt{ANDROID\_SOCKET\_}'' prefix, the rest of the environment variable name is saved as a RESERVED socket address.

\subsection{Peer Credential Check Extraction}
\label{peercredextraction}

The \texttt{getsockopt} system call~\cite{getsockopt}, when invoked with a file descriptor \texttt{sockfd}, retrieves the value of various options for the socket pointed to by \texttt{sockfd}. The option retrieved is specified by the \texttt{optname} argument, which is an integer corresponding to a valid socket option. The retrieved option is stored in the pointer specified by the \texttt{optval} argument. In our case, we are mainly interested in \texttt{getsockopt} calls where the \texttt{optname} is specified as \texttt{SO\_PEERCRED} (0x11). In this case, the connected peer's credentials are stored at the \texttt{optval} pointer in a \texttt{ucred} struct. This struct contains three member variables: the process ID (PID), user ID (UID) and group ID (GID) of the connected peer. 

Using the same CFG used in the Socket Address Extraction step, we perform another dataflow analysis of every function that invokes the \texttt{getsockopt} system call. First, we check the value of the option name (\texttt{optname}) argument. If the function is called with the \texttt{SO\_PEERCRED} optname, we track all subsequent uses of the returned credentials in the function and record which credentials are being used. We use the same categorization used in~\cite{uds}; UID- and GID-based checks are considered secure, while PID checks are considered weak. 
Additionally, we attempt to detect and categorize uses of these credentials. We have identified two types of uses:
(1) \textit{Integer comparisons: } We detect whenever a credential is used in a comparison instruction and record the operand if it is a constant integer, or ``UNDEFINED'' if it is not.
(2) \textit{Function arguments: } We detect whenever a credential is used as a function argument and record the function address and name (if it was not stripped).
With this usage information, we can determine exactly what credentials a connected peer needs to be able to communicate with the process through a given socket, thus greatly aiding in further interpretation of the analysis result. 


\subsection{File Permission Analysis}

Following the detection of all socket addresses and their filesystem permissions (if any), we check whether an untrusted app with all possible permission-mapped GIDs can access the socket file for each FILESYSTEM or RESERVED socket address. We use a modified version of \BigMAC{} to reason about the MAC policy and determine whether an untrusted app has access to a socket file. We also inspect the socket file's permission bits, UID and GID to determine access w.r.t.\ the DAC policy. 

\section{Implementation}

In this section, we discuss the implementation of the three most crucial parts of the \ourtool~architecture, which includes the extension of \BigMAC{} to better serve our use case, the static analysis of daemon binaries using \texttt{angr}~\cite{angr}, and the final step of filtering and categorizing accessible sockets.

\subsection{BigMAC Query}

The first step following successful extraction of a firmware image is the SELinux policy analysis. We implemented an easy-to-use API that exposes the most crucial functionalities of \BigMAC{} to the developer. This API facilitates running the whole workflow of \BigMAC{} in a single call, and implements an interface to the prologue engine that facilitates query operations on the generated Attack Graph. We use this API by specifying the path of the extracted SELinux policy and running the attack graph instantiation. Using this attack graph, we run the query: \texttt{query(untrusted\_app, \_, 1)} to retrieve all nodes in the graph that an untrusted app can write/connect to. We then filter only socket objects from the resultant list. Since socket objects are \texttt{IPCNodes}, they hold a reference to the owning process. Thus, we can retrieve all the processes, and their executable binaries, that an untrusted app can connect to through a Unix domain socket. 

Additionally, we extended \BigMAC{}'s init boot simulation step to handle \texttt{socket} options under service definitions in init RC files. On encountering a \texttt{socket} entry under a service definition, \BigMAC{} now creates the corresponding file in the simulated filesystem as part of the boot process with the specified permissions. Additionally, it assigns the correct SELinux context to the socket file based on the extracted filesystem contexts. This addition is essential as \BigMAC{} removes filesystem contexts that do not have a backing file from the attack graph which would have prevented us from querying whether an untrusted app has access to these socket files.

\begin{table}[t]
\caption{Android-specific bind APIs}
\centering
\label{table:bind_apis}
\begin{tabular}{|l|l|l|}
\hline
\textbf{Function}             & \textbf{Namespace} & \textbf{Library} \\ \hline
FrameworkListener             & RESERVED           & libsysutils.so   \\ \hline
SocketListener                & RESERVED           & libsysutils.so   \\ \hline
android\_get\_control\_socket & RESERVED           & libcutils.so     \\ \hline
socket\_local\_server\_bind   & Any                & libcutils.so     \\ \hline
socket\_local\_server         & Any                & libcutils.so     \\ \hline
\end{tabular}
\end{table}

\subsection{Static Binary Analysis}

Our static binary analysis implementation contains three modules (about 2K LoC).  The Socket Address Extraction module extracts socket addresses that the binary is listening over by analyzing \texttt{bind} call sites. The DAC Check Extraction module detects and analyzes DAC checks by performing data flow analysis after \texttt{getsockopt} calls with the \texttt{SO\_PEERCRED} argument as discussed in Section~\ref{peercredextraction}. Each daemon binary an untrusted app can connect to is statically analyzed to retrieve all the Unix domain socket addresses it listens on and the permissions they are created with, as well as detect any hardcoded DAC checks in the binary. 

\begin{sloppypar}

We implement our static binary analysis using the \texttt{angr} framework~\cite{angr}. We first generate the CFG of the analyzed binary during which function prologues are detected and stored in the \texttt{angr} project's knowledge base. The dataflow analysis is implemented by \texttt{angr}'s intra-procedural \texttt{ReachingDefinitions} analysis~\cite{rda} and is used in both socket address extraction and DAC check extraction. To make the analysis inter-procedural, we implement a \texttt{FunctionHandler} which handles function calls by performing the \texttt{ReachingDefinitions} analysis recursively based on~\cite{rda_ip}.

\end{sloppypar}

\subhead{Socket Address Extraction}
First, we find all call sites to the \texttt{bind} system call. This is done by finding the \texttt{bind} function node in the CFG and listing all of its predecessor nodes where the connecting edge is of type \texttt{Ijk\_Call}, signifying a function call. For each one of these nodes, we find the function that it belongs to, and perform an inter-procedural data flow analysis on that function. The \texttt{FunctionHandler} also simulates common libc string manipulation functions, such as \texttt{strcpy}, \texttt{sprintf} and others, in order to capture dynamically constructed socket addresses at the \texttt{bind} callsite. These string manipulation handlers are implemented in order to avoid inaccuracies caused by the complex control flow structures associated with string operations. Additionally, Android provides additional utility APIs for system daemons to create, bind, and listen over local sockets. These functions are found in \texttt{libcutils.so} and \texttt{libsysutils.so} and are detailed in Table \ref{table:bind_apis}. We perform the same analysis at the call sites of these functions to recover socket addresses passed to these utility functions.


\subhead{DAC Check Extraction}
The same dataflow analysis implementation is used for DAC check extraction. We analyze call sites of the \texttt{getsockopt} system call, and check its arguments. If the \texttt{optname} argument is set to \texttt{SO\_PEERCRED}, we track usages of the \texttt{ucred} struct, stored in the pointer \texttt{optval}, by tainting its member variables. We record any variables used and attempt to identify the type of usage as discussed in Section~\ref{peercredextraction}.

\subsection{File Permission Analysis}
We add additional functionality to \BigMAC{} allowing us to add previously undetected files, e.g., files created dynamically by a running process. Following the recovery of FILESYSTEM socket addresses and their DAC metadata, we insert these filenames along with their metadata into \BigMAC{}'s recovered filesystem, and rerun \BigMAC{}'s workflow. This will assign the correct SELinux labels to these files in an automated manner, which allows us to determine whether a socket file is accessible through simple queries of the Prolog engine.


\begin{figure}[t!]
     \centering
     \includegraphics[width=0.48\textwidth]{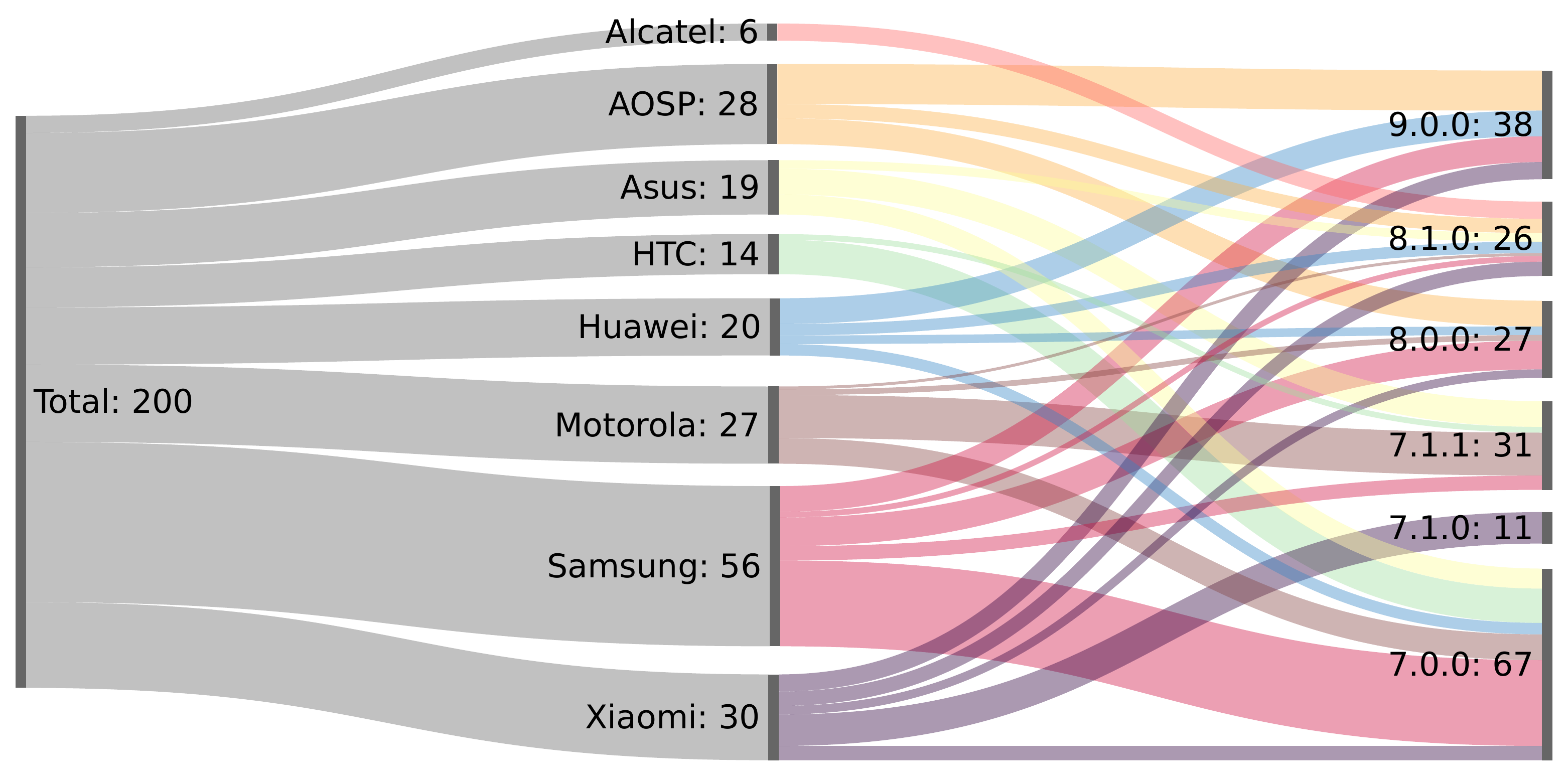}
     \caption{Distribution of Android versions and vendors in our corpus}
     \label{fig:firmware_images}
\end{figure}

\section{Evaluation}

In this section, we first present our findings on accessible sockets, and security downgrade; we then report on the performance measurements and ground truth comparison. The number of firmware images analyzed per vendor and Android version can be found in Figure~\ref{fig:firmware_images}. Accessible sockets are the sockets that we have identified to be accessible for an untrusted app without any prerequisites (following our threat model). For security downgrade, we consider daemons that exist in AOSP Android but have weaker security protections due to vendor customization of access control policies and customization of the daemons themselves.

\begin{table*}[]
\caption{Socket addresses an untrusted app can connect to, their system daemons, 
authentication checks they implement and the vendors where they were found to be accessible }
\label{table:accessible_sockets}
\centering
\begin{tabular}{|l|l|l|l|l|}
\hline
\textbf{Address} & \textbf{Namespace} & \textbf{System Daemon}       & \textbf{Auth.\ Checks}                                 & \textbf{Vendor}             \\ \hline
/dev/socket/nims              & FILESYSTEM & cnd        & None                         & Asus, HTC, Motorola, Xiaomi \\ \hline
cnd              & RESERVED           & cnd               & 
GID or AppName & Asus, HTC, Motorola, Xiaomi \\ \hline
qvrservice       & RESERVED           & qvrservice & None                                                           & Asus, HTC, Samsung, Xiaomi  \\ \hline
qvrservice\_camera       & RESERVED           & qvrservice & None                                                           & Xiaomi  \\ \hline
seempdw                       & RESERVED   & seempd     & None                         & Asus, HTC, Xiaomi           \\ \hline
@fmhal\_sock     & ABSTRACT           & fmhal\_service    & UID                                   & Asus, Motorola, Xiaomi      \\ \hline
@qcom.dun.server              & ABSTRACT   & dun-server & None                         & Asus, Xiaomi                \\ \hline
@cand.socket.ctrl             & ABSTRACT   & cand       & None                         & HTC                         \\ \hline
@cand.socket.msg              & ABSTRACT   & cand       & None                         & HTC                         \\ \hline
dmagent                       & RESERVED   & dmagent    & None                         & HTC                         \\ \hline
cfiat                         & RESERVED   & rild       & UID & HTC                         \\ \hline
kipc                          & RESERVED   & rild       & UID & HTC                         \\ \hline
/dev/socket/dpmwrapper        & FILESYSTEM & dpmd       & None                         & HTC, Xiaomi                 \\ \hline
@btloggersock                 & ABSTRACT   & bt\_logger & None                         & Motorola, Xiaomi            \\ \hline
@dev/socket/jack/set.priority & ABSTRACT   & apaservice & None                         & Samsung                     \\ \hline
napproxyd                     & RESERVED   & netd       & None                         & Samsung                     \\ \hline
tcm                           & RESERVED   & dpmd       & None                         & Xiaomi, Asus                      \\ \hline
dpmwrapper                    & RESERVED   & dpmd       & None                         & Xiaomi, Asus                     \\ \hline
\end{tabular}
\end{table*}

\begin{figure}[t]
    \centering
    \includegraphics[width=0.5\textwidth]{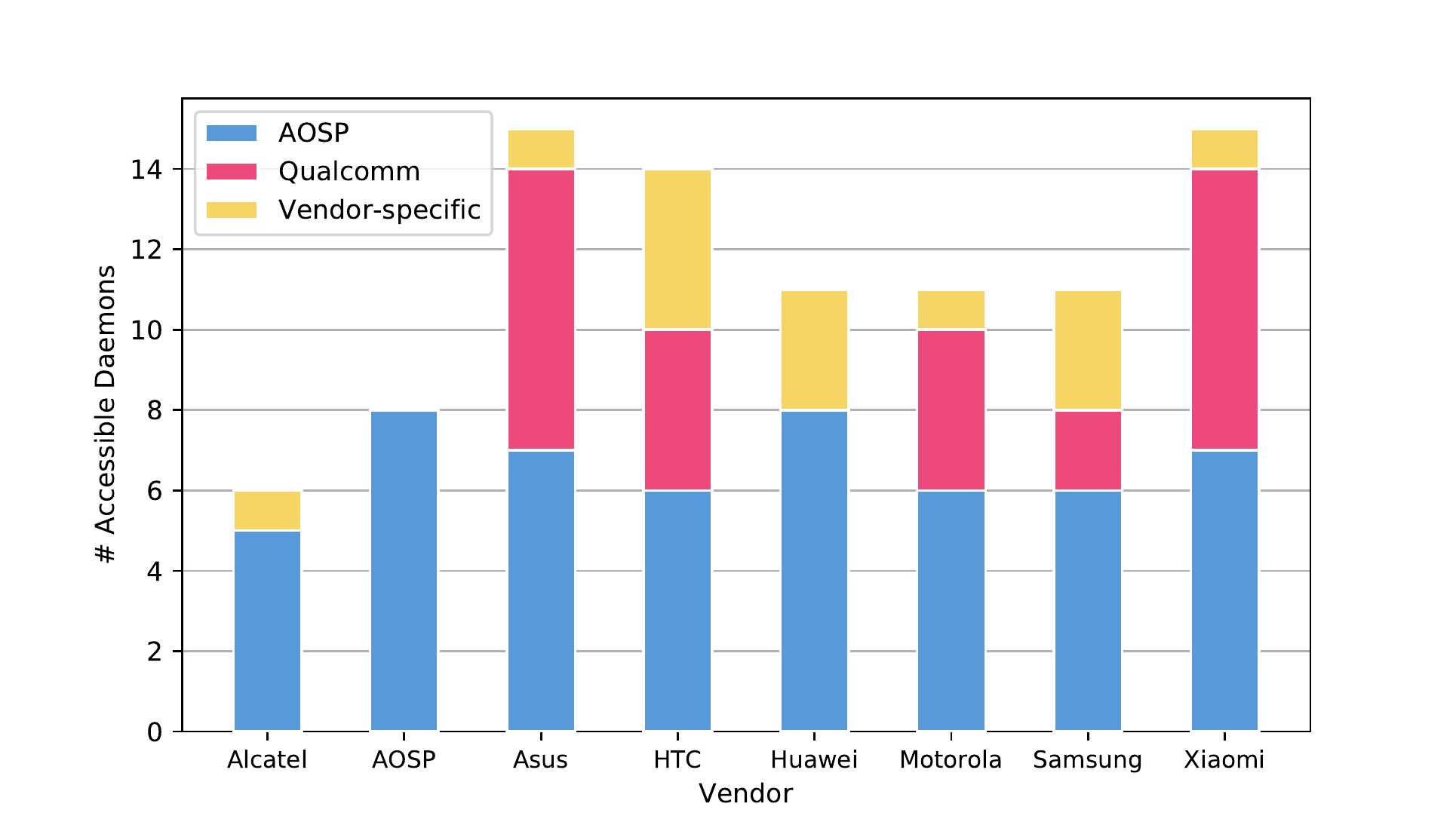}
    \caption{Accessible system daemons by vendor in our corpus}
    \label{fig:vendor_specific}
\end{figure}

\label{sec:accessible_sockets}
\subsection{Accessible Sockets}

Figure~\ref{fig:vendor_specific} displays the number of accessible system daemons by vendor. In this figure, we count daemons that are accessible to an untrusted app by SELinux and for which we have found at least one socket. We divide them into three categories: AOSP daemons, Qualcomm daemons and vendor-specific daemons. We count AOSP daemons with additional vendor-specific sockets as vendor-specific daemons. From these daemons, we identified 28 unique socket addresses that an untrusted app is allowed to connect to by the access control policy in at least one of the firmware images analyzed.
We present 17 of these socket addresses and their daemons in Table~\ref{table:accessible_sockets}. We omit the remaining 11 sockets from the table as they are intended to be accessible to untrusted apps as part of the Android framework. We also specify any authentication checks performed 
on the client after a connection is established, as well as the list of vendors these socket addresses were detected on.\footnote{We only specify the vendors with firmware images where the system daemon that uses these sockets is \emph{accessible}. If the daemon exists in other vendors' firmware images, we do not include it.} In the following, we discuss each daemon's functionality and its accessible sockets. Information about proprietary daemons' functionality is not publicly available. Therefore, we infer their functionality based on static analysis of the binaries, as well as whatever information we can find online. 

\subsubsection{AOSP Daemons} AOSP daemons are available in all Android systems as part of AOSP 
Android. The AOSP version of these daemons' source code is made available by AOSP, although vendors might make proprietary customizations to them in their own distributions of Android. These daemons purposefully expose sockets for communication with an untrusted app to implement various functionalities. These sockets were detected consistently across our dataset, confirming our tool's reliability at detecting accessible sockets. Since these sockets are intended to be accessed by untrusted apps, we omit these accessible sockets from Table~\ref{table:accessible_sockets} and briefly discuss the daemons' functionality instead. 


The \textit{logd} daemon is a centralized logger implementing all logging operations in Android~\cite{daemonsBook}. It utilizes three sockets, all of which are accessible to an untrusted app: logd, logdr, and logdw. The \textit{netd} daemon is responsible for managing network interface configurations~\cite{daemonsBook}. In AOSP Android, the \textit{netd} daemon utilizes four socket addresses: netd, dnsproxyd, mdns and fwmarkd. Of these sockets, dnsproxyd and fwmarkd are accessible to an untrusted app with the \texttt{INTERNET} permission. The \textit{surfaceflinger} daemon's main functionality is to compose and render multiple display surfaces onto the display~\cite{daemonsBook}. In Android 8.0+ images, we found three accessible RESERVED socket addresses from the \textit{surfaceflinger} daemon binary located in the pdx/system/vr/display/ directory and include client, manager, and vsync. 
The \textit{tombstoned} daemon was added in Android 8.0 and it plays a role in capturing crash data from a system and storing it for further analysis. The \textit{traced} daemon is part of an open-source solution developed by Perfetto~\cite{perfetto} and used in Android for system profiling, app tracing and trace analysis. The \textit{perfd} daemon collects information to keep track of performance on the system. 

\subsubsection{Qualcomm Daemons} Qualcomm provides a wide range of  hardware and peripherals on Android devices, such as the processor and the Mobile Station on Modem (MSM) system on chip (SoC). For interoperability between these peripherals and the operating system, Qualcomm implements daemons that bridge the communication between these devices and the rest of the Android framework. These daemons were found to be accessible across multiple vendors' images in our dataset. In AOSP however, these daemons exist, but none of them are accessible to an untrusted app. This discrepancy 
indicates that access control policies placed by AOSP were relaxed by other vendors where these daemons were found to be accessible.


 \subhead{cnd} The \textit{cnd} daemon manages Qualcomm Connectivity Engine which chooses between 3G/4G and Wi-Fi networks based on their performance for the specific application a user is using~\cite{cnd}. It is a proprietary daemon, therefore its exact functionality is not publicly known. In Android versions prior to 8.0, it uses the cnd RESERVED socket and a FILESYSTEM socket located in /dev/socket/nims, both of which are accessible to an untrusted app with the \texttt{INTERNET} permission. The /dev/socket/nims socket was also found to be world-accessible in some HTC, Motorola and Xiaomi images. It is unclear whether this is a result of misconfiguration or a change of the socket's functionality.

 \subhead{qvrservice} The \textit{qvrservice} daemon is a proprietary daemon that manages Qualcomm VR service. It exposes a world-accessible RESERVED socket qvrservice. On Xiaomi Android 9.0 images, it also exposes the qvr\_camera RESERVED socket with the same permissions.

 \subhead{seempd} The \textit{seempd} daemon is part of the Qualcomm Trusted Execution Environment stack (QTEE~\cite{qtee}). It exposes a world-writable DGRAM socket with address seempdw.

 \subhead{dpmd} The \textit{dpmd} is a daemon which stands for Data Port Mapper, and is part of the QTI DPM Framework~\cite{dpm}. It is unclear what functionality it provides. This daemon utilizes two sockets: a RESERVED socket with address dpmd and a FILESYSTEM socket with address /dev/socket/dpmwrapper. The dpmd socket is configured to be inaccessible to apps. In Android 8.0 images, it uses an additional RESERVED socket named tcm, and the FILESYSTEM socket /dev/socket/dpmwrapper was changed to a RESERVED socket ``dpmwrapper''. The dpmwrapper and tcm RESERVED sockets require the \texttt{INTERNET} permission to be able to connect.

 \subhead{dun-server} \textit{dun-server} is a daemon that implements and manages Dial-Up Networking over Bluetooth~\cite{dun-server}. It listens over the @qcom.dun.server ABSTRACT socket address. It contains no authentication check, allowing any client to connect to dun-server over this socket. Additionally, manual static analysis of the binary reveals that this socket is bound and closed repeatedly by \textit{dun-server}. Since this is an ABSTRACT socket, any process can create one with the same name as long as that address is not being used. Therefore, a malicious process can bind to this socket address before \textit{dun-server} re-binds it, denying \textit{dun-server} from using it and disrupting its workflow.


 \subhead{fmhal\_service} \textit{fmhal\_service} is an open-source daemon that manages FM radio on supported systems~\cite{fmhal_service}. It exposes an ABSTRACT socket with address @fmhal\_sock. Since this socket is ABSTRACT, it is accessible by default. Thus, any app can establish a connection to it. However, when a client connects, a UID check is performed to ensure that the client's UID is one of root, system or bluetooth.

 \subhead{bt\_logger} \textit{bt\_logger} is an open-source daemon that has the ability to log Bluetooth traffic~\cite{bt_logger}. It exposes an ABSTRACT socket with address @btloggersock with no authentication check, allowing any client to start/stop Bluetooth snooping. 

\subsubsection{Vendor-specific Daemons}\label{sec:vendor_specific} Vendor-specific daemons are daemons that are developed by the Android device manufacturer and bundled with their operating system distribution. A daemon is classified as vendor-specific if it is present in the Android images of a single vendor. In our results, two out of three accessible vendor-specific daemons run as root, presenting valuable targets for exploitation. The third daemon provides an interface to a function available only to processes running privileged UIDs.


 \subhead{dmagent} HTC \textit{dmagent} is a proprietary daemon that manages the Open Media Alliance Device Management protocol. dmagent runs with UID root and listens over the RESERVED socket address dmagent. The socket is configured to be accessible to applications with the \texttt{INTERNET} permission. This socket has been previously used to issue copy file command to the daemon which acts to copy files as root from arbitrary source to arbitrary destination~\cite{weaksauce}. Our analysis shows that this socket remains unprotected by access control policies or authentication checks, making it a prime target for exploitation of a root process. 
 
 \subhead{cand} HTC \textit{cand} is a proprietary daemon which runs as root and listens over two ABSTRACT socket addresses: @cand.socket.ctrl and @cand.socket.msg. Through static analysis, we determined that the daemon  serves as an interface to communicate over the CANBus, although the specific use case is not clear. Nevertheless, since it runs as root and listens over unprotected ABSTRACT sockets, it may present another security risk much like dmagent.

 \subhead{apaservice} The \textit{apaservice} daemon is part of the Samsung Android Professional Audio framework~\cite{apaservice}. It runs with a UID of ``jack.'' We found that it creates and listens over one ABSTRACT socket address @dev/socket/jack/set.priority which is used as an interface through which it calls \texttt{android::requestPriority} with the parameters it receives. According to AOSP source code~\cite{schedpolsvc}, this functionality should only be exposed to processes with the audioserver, cameraserver and bluetooth UIDs. 


\subsection{Downgraded Security}

\label{sec:downgraded_security}

A system daemon is considered to have downgraded security if the vendor relaxes SELinux rules that would have prevented communication between the daemon and an untrusted app in AOSP. To find these instances of downgraded security, we go over the list of daemons an untrusted app can communicate with for each vendor. We then try to find each daemon in that list and its service definition in the corresponding AOSP Android version. If the service exists in AOSP and is enabled, but is not accessible by an untrusted app, then we flag it as a security downgrade. 

\subhead{HTC dumpstate} The \textit{dumpstate} system daemon is an AOSP system daemon that can generate logs that are used to collect details of device-specific issues; an untrusted app is disallowed to communicate with this daemon in AOSP. HTC relaxed this restriction and added two extra sockets to the daemon: htc\_dk and htc\_dlk. Untrusted app access to these sockets is not allowed by both the MAC and DAC policies. However, as per the comments of the \texttt{file\_contexts} file, the htc\_dlk socket sends kernel log messages to a system app. This is a bad security practice as kernel logs can hold sensitive information, and pre-installed apps packaged with vendor-customized firmware have been shown to be insecure~\cite{firmscope}.

\subhead{HTC rild} \textit{rild} is the Radio Interface Layer daemon in Android~\cite{RadioLay44:online, rild}. It provides an abstraction layer between Android telephony services layer and the radio hardware layer and handles all telephony operations such as call handling, SMS, and others. In Android versions prior to 8.0, \textit{rild} utilizes three sockets: rild, rild-debug and sap\_uim\_socket1. In AOSP Android, communication with rild using Unix domain sockets is not allowed for untrusted apps by the SELinux policy. In HTC images, our SELinux policy analysis showed that the policy was relaxed and an untrusted app was allowed to communicate with \textit{rild}. Furthermore, we detected two vendor-specific sockets, kipc and cfiat, both of which grant read and write access to an untrusted app with the \texttt{INTERNET} permission. These socket addresses have only been detected on the HTC firmware images we analyzed, and their file contexts are labeled htc\_cfiat\_socket and htc\_kipc\_socket, confirming that they originate from an HTC-specific vendor customization of \textit{rild}. We detected a UID-based authentication check in the HTC \textit{rild} binary, leaving these sockets potentially protected only by a single post-connection DAC check. Therefore, if a malicious app changes its UID through a privilege escalation exploit, it can gain access to these sockets, which would not have been possible if the SELinux policy had not been relaxed. 

\subhead{cnd} The \textit{cnd} daemon can be available in AOSP firmware images. In all of the AOSP and Samsung images tested, an untrusted app does not have access to this daemon. On the other hand, Asus, HTC, Motorola and Xiaomi firmware images allow an untrusted app to communicate with this daemon. In this case, the daemon exposes two sockets: cnd and /dev/socket/nims that are accessible to an untrusted app, one of which has no authentication checks. Both require the app to have the \texttt{INTERNET} permission.

\begin{sloppypar}

\subhead{Asus mm-qcamera-daemon} In Asus Android images, \textit{mm-qcamera-daemon} is allowed to communicate with an untrusted app. It contains a socket named at address /data/misc/camera/cam\_socket in the FILESYSTEM namespace. The socket itself is inaccessible by both the MAC and DAC policies.
\end{sloppypar}

\subsection{Abstract Sockets}
Our analysis aims to find sockets accessible to untrusted apps. However, we report on the ABSTRACT sockets in our results that are vulnerable to DoS attacks by untrusted apps. Such sockets become vulnerable to DoS if the daemon that owns the socket closes the socket at any point in its operation, or if the daemon exits for any reason. We detect the first case through manual static analysis using Ghidra~\cite{ghidra}. This is done by detecting \texttt{close} calls on the socket that was previously bound to an ABSTRACT address. If \texttt{close} is called anywhere outside of a final cleanup of the daemon's resources during termination, then we flag it as vulnerable to DoS. As for the second case, we detect daemons that are started or stopped by property triggers. These triggers are defined in init RC files and are used to start/stop a service daemon when the specified system property changes, depending on the value of the system property. A malicious app can cause DoS of the vulnerable daemon in either of these cases by repeatedly attempting to bind the ABSTRACT socket that the daemon would usually bind. As a result, if the daemon is stopped, restarted, or closes the ABSTRACT socket, the malicious app will be successful in binding it. Consequently, the daemon will fail to bind this ABSTRACT socket again, disrupting IPC between the daemon and other processes that rely on it.


Four of the ABSTRACT sockets that we detected are vulnerable to DoS, as they match the criteria defined above: @dev/socket/jack/set.priority, @fmhal\_service, @qcom.dun.server and @btloggersock. @dev/socket/jack/set.priority is bound by \textit{apaservice} and is triggered by a service call to \texttt{IAPAService::StartJackd}. Therefore, a malicious app can occupy this ABSTRACT socket address before another app makes the service call to \texttt{IAPAService::StartJackd}. When the service call is made, \textit{apaservice} would fail to bind the socket. However, this failure is handled gracefully by \textit{apaservice} so that its other functionalities would remain unaffected. @fmhal\_service and @btloggersock are bound by their system daemons on initialization, but the daemons themselves are triggered by a property trigger in init RC files. A malicious app can bind either of these addresses while the corresponding property is not set. When the property is set and the daemons are started, they fail to bind the socket address and terminate due to the resulting bind error. @qcom.dun.server is bound and closed repeatedly by \textit{dun-server} after every connection. Exploiting this DoS vulnerability would require a race-condition, where a malicious app attempts to bind the @qcom.dun.server address before \textit{dun-server} re-binds it.



Additionally, by applying our framework to find accessible sockets for the \texttt{system\_app} context, we found two ABSTRACT sockets, @com.mtk.aee.aed and @aee:rttd created by the \textit{aee\_aed} daemon in the Alcatel Android 8.1 firmware. AEE stands for Android Exception Enhancement, and the daemon serves to collect and log backtraces of crashes on the system. We found that the daemon restarts on any configuration change, e.g.,  triggered by a system app, allowing a malicious app to occupy this ABSTRACT socket address and resulting in DoS of the daemon. This can stop the daemon from logging crash information on the system, potentially hiding evidence of exploit attempts.


\subsection{Ground Truth Evaluation}
To confirm the correctness of our framework in detecting sockets and their access control properties, we ran a ground truth evaluation on Samsung devices running Android 7.0 and 8.0, and a Motorola device running Android 7.1.1. The test was carried out by an app we developed which obtains the permissions we listed in our threat model. The app runs the \texttt{netstat -xl} command to list all the listening Unix domain sockets and their addresses. The app then tries to connect to each socket address and displays a table containing the socket addresses and the result of the connection attempt. However, this app is not part of our framework and only serves to collect ground truth data for evaluation.

On all devices, the app successfully connected to the fwmarkd, dnsproxyd, logd, logdr RESERVED socket addresses. On the Motorola device, the app also reported a connection to the perfd RESERVED socket address. All of these socket addresses were accurately detected by our framework as accessible sockets. For the Samsung device, our analysis detected an ABSTRACT socket owned by the \textit{apaservice} daemon at @dev/socket/jack/set.priority which was not found on the running device. We later discover that the socket is created only after a Samsung-specific service is activated through a Binder call and discuss the details in Section~\ref{sec:samsung_socket}. This demonstrates the effectiveness of our approach compared to dynamic analysis which may not detect sockets created conditionally or in response to a trigger. Note that our ground truth evaluation app is a simplified implementation of the \textit{Connection Tester} dynamic analysis module used in~\cite{uds}. Thus, we claim that our approach achieves a better socket detection rate due to the higher coverage inherent to static analysis. The accessible sockets detected by the testing app can be found in Table~\ref{table:gte} in Appendix~\ref{appendix:ground_truth}.


\subsection{Performance Evaluation}

\begin{figure}[t]
    \centering
    \includegraphics[width=0.5\textwidth]{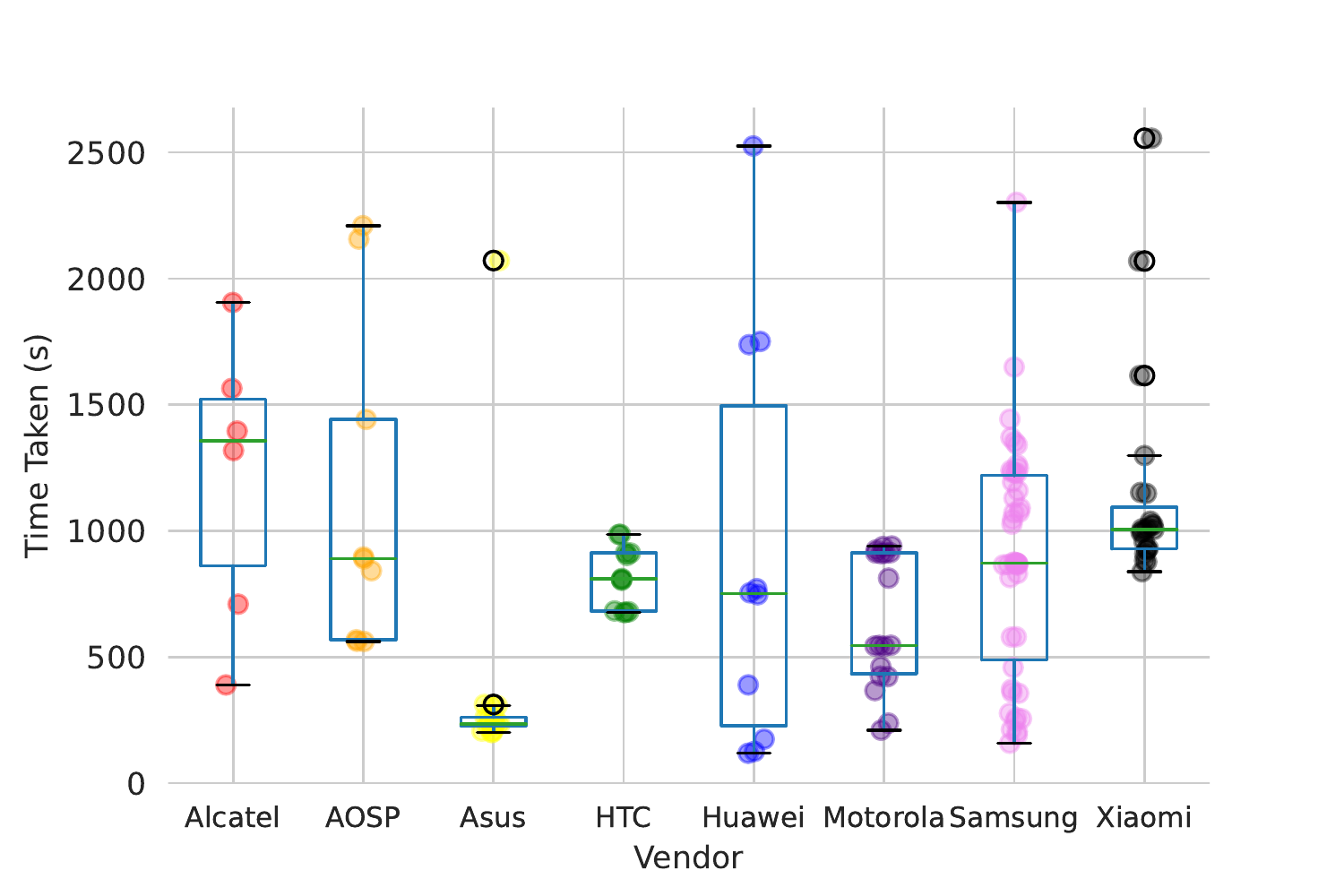}
    \caption{Analysis time for firmware images in our corpus}
    \label{fig:performance}
\end{figure}

We ran our analysis on a PC with the Intel(R) Core(TM) i7-8700 CPU @3.20GHz and 16 GB RAM. On average, each firmware was analyzed in 770 seconds (approx.\ 14 minutes). Figure~\ref{fig:performance} shows a box plot of the time taken for each firmware image. The static binary analysis takes a majority of that time at an average of 736 seconds (approx.\ 14 minutes), resulting in a large variance for each image, depending on how many binaries are being analyzed, i.e., how many service daemons are accessible to an untrusted app, and how complex their binaries are. The remaining time is used in the initial and final steps in the instantiation and querying of \BigMAC{}. Currently, our prototype does not implement obvious optimizations, such as parallelization of the binary analysis step for each binary. We leave these engineering tasks as future improvements to our framework. Based on monitoring the memory usage for our analysis, parallelization would have resulted in a 3x speedup on the same PC.

\section{Case Studies and Responsible Disclosure}

\label{sec:samsung_socket}

In this section, we focus on two interesting cases from our results that translate to vulnerabilities that a malicious app can potentially exploit. 
We discuss Samsung \textit{apaservice} daemon's FILESYSTEM socket and how it can be used to request a priority for any process ID or thread ID, a functionality only available to processes with the audioserver, cameraserver or bluetooth UID. We discuss another potential permission bypass exploit in Qualcomm's \textit{cnd} and \textit{dpmd} daemons, both of which implement an identical check based on the connecting process's name. 





\begin{sloppypar}
\subhead{Samsung apaservice}
Our analysis indicates that Samsung's \textit{apaservice} daemon creates an ABSTRACT socket with address @dev/socket/jack/set.priority. In our ground truth evaluation, this socket was not found at first on the running device. We analyzed the \textit{apaservice} binary to determine the reason, and found that this socket is only created after calling \texttt{APAService::startJackd.} This method is exposed by the service \texttt{com.samsung.android.jam.IAPAService.} After calling this method using the \texttt{service call} command, the socket was created and appeared in the \texttt{netstat} output. This demonstrates the effectiveness of our analysis in uncovering sockets which are created only under certain conditions, as compared to dynamic analysis. 

The socket at @dev/socket/jack/set.priority is a DGRAM socket under \textit{apaservice} that accepts messages from any client, with no DAC check after receiving a message. It receives messages of the format ``\texttt{*4<pid>,<tid>,<priority>},'' where \texttt{<pid>} is a process ID, \texttt{tid} is a thread ID, and \texttt{<priority>} is the requested priority. These values are then passed to the function \texttt{android::requestPriority}, which requests the SchedulingPolicyService to assign a priority to the requested process ID and thread ID. Although this request is typically available to system processes running under the audioserver, cameraserver and bluetooth UIDs, an untrusted app can set its own priority, or the priority of any other process through this socket, effectively bypassing authentication checks.

Additionally, through manual analysis, we discovered that the function that handles messages received over this socket, \texttt{android::APAService::} \texttt{handlePriorityMessage}, is vulnerable to buffer overflow. By sending the correct preamble, ``4*'', followed by 25 bytes of data, the \textit{apaservice} daemon crashes due to stack corruption. The backtrace logs show that the return address was successfully overwritten. The impact of this buffer overflow vulnerability can range from DoS of \textit{apaservice} to privilege escalation to the more privileged ``jack'' UID (recall that this UID is used by \textit{apaservice}) in the less restrictive \texttt{apaservice} SELinux context. We developed a Proof-of-Concept (PoC) for this buffer overflow that crashes the daemon, achieving DoS. Achieving local code execution would require bypassing the buffer overflow protections compiled into the daemon binary. \textit{Apaservice} exists in Samsung Android up to version 8.1. Within our analysis, we first noted \textit{apaservice} in Samsung Android 7.0. This indicates the lifetime of this vulnerability would, at least, be from the release of Samsung Android 7.0 on Aug.\ 22, 2016 through when the vulnerability was reported on Sept.\ 10, 2021 by us. 

\end{sloppypar}

\subhead{Qualcomm cnd and dpmd}
In 18 Xiaomi and four HTC images analyzed, we found that \textit{dpmd} implements an insecure authentication check after a connection is established. The \textit{cnd} daemon implements an identical authentication check in 14 HTC images and seven Motorola images. Both of these daemons are started by init with root UID, but later drop to system UID through a \texttt{setuid} call. The authentication mechanism works as follows: after a connection with a new client is established, the client's DAC credentials are retrieved using a \texttt{getsockopt} call. First, the GID is compared against a list of GIDs that are allowed by the daemon. If the GID does not match any of the allowed GIDs, then the PID is used to retrieve the connecting client's process name by reading the \texttt{/proc/<pid>/comm} file for that process. In Unix systems, this file exists for every process and contains the process name. The process name is then compared to a list of allowed process names, and access is granted if a match is found. This is however an insecure check, as any app can change its own process name dynamically, even if a different process has the same name. Therefore, a malicious app can bypass this check trivially by changing its name to that of an allowed process. 

In the case of \textit{dpmd}, this check is implemented for an inaccessible RESERVED socket of the same name, ``dpmd,'' and an untrusted app is not allowed to connect to by both MAC and DAC. On the other hand, \textit{cnd} implements this check on its ``cnd'' RESERVED socket, which is accessible to an untrusted app. Through static analysis, we infer that this socket allows clients to get/set network settings such as WiFiAP, WiFi P2P, and Default Network settings, by sending the appropriate command over the cnd socket. 

\label{sec:responsible_disclosure}
\subhead{Responsible Disclosure} We sent detailed reports of these vulnerabilities to Samsung and Qualcomm, including the aforementioned PoC for Samsung. Samsung acknowledged the vulnerability in \textit{apaservice} and patched it in in SMR Sep-2021 Release 1. The vulnerability was assigned CVE-2021-25461. They also rewarded our findings through their bug bounty program on Bugcrowd. Qualcomm's response to our disclosure was that the unprotected \textit{cnd} socket is deprecated starting Android 8.0. Thus, they will not be patching the affected systems, despite around 180 million users relying on the affected Android versions (7.0-7.1)~\cite{statcounter_versions}. As for \textit{dpmd} daemon, they mention that the daemon now uses a more secure UID check although we still see the same issue up to Android 9.0 firmware in our dataset.

\section{Limitations}

Our analysis approach faces certain limitations. Firmware unpacking and extraction presents the only obstacle to expanding our analysis to more recent Android versions and a wider variety of vendors. Extending the current open-source toolset for Android image extraction requires significant engineering effort, but can pave the way for similar large-scale analyses. Additionally, we discuss the limitations inherent in our static binary analysis approach that make the analysis of statically-linked stripped binaries difficult.

\subhead{Firmware Unpacking and Extraction}
Extracting and unpacking Android images is not trivial as the format of factory images can vary greatly between different Android vendors and versions. Multiple tools have been developed that facilitate the unpacking process or different stages of it. However, to our knowledge, there is no freely available unified factory image unpacking tool that can unpack any firmware image across different versions and vendors, except for the one we used~\cite{at_commands, trustbutverify}. These tools are outdated, however, and only support Android versions 5-9 for AOSP, and 5-8 for other vendors. Furthermore, within these versions the unpacking success rate is not perfect, and some filesystems may not be recovered. This limits the operable dataset we can use in our analysis, and as a result extracted firmware might have missing daemons or SELinux policy files. 



\subhead{Static Binary Analysis}
Our implementation of the static binary analysis relies on detecting Android bind APIs and string manipulation functions by their symbol name. This does not pose a problem in the case of dynamically-linked binaries 
since external symbols are persevered for linking. However, this becomes problematic in statically-linked stripped binaries. In our analysis, we encountered three cases of statically-linked stripped binaries which we ultimately skipped, namely: \textit{mcDriverDaemon}, \textit{debuggerd} and \textit{adbd}. Additionally, we assume that if a \texttt{bind} call exists in the binary with a Unix domain socket address parameter, then that socket is bound on initialization of the daemon. We do not perform reachability analysis to avoid the problem of inaccurately resolving indirect jumps. Additionally, \ourtool{}'s implementation contains over-approximations that could 
 produce similar duplicate results (e.g. \texttt{@dev/socket/some\_socket} vs.\ \texttt{/dev/socket/some\_socket}) in some cases where the socket address is constructed in multiple string manipulation steps.

\subhead{Manual Analysis} 
\ourtool's output is a report of all socket addresses detected by our analysis, and any DAC checks or file permissions assigned to them. A manual reverse engineering/analysis process needs to be carried out in order to evaluate whether these sockets result in a vulnerability. For instance, in the case of ABSTRACT sockets, the analyst would examine the binary's code to determine if the socket is closed and re-bound in a loop, which would lead to the socket being vulnerable to DoS. This is a limitation since it requires manual effort to probe the results for vulnerabilities.


\section{Conclusion}
In this work, we present \ourtool, a static analysis framework to evaluate the security of Unix domain sockets used in Android. Our approach combines fine-grained access control policy analysis with static binary analysis techniques to comprehensively detect exposed IPC sockets available to an untrusted app. We use this framework to analyze \datasetsize{} Android images from different vendors and Android versions, and uncover vulnerabilities and access control misconfigurations, such as permission bypass and denial of service. Some of these sockets would not have been discovered by previous work relying on dynamic analysis. We will open-source our static binary analysis module to make it available for the community upon publishing. 
The source code for our static binary analysis module is published at https://github.com/mounir-khaled/SAUSAGE.


\subsection*{Acknowledgements}
This work was supported in part by the Office of Naval Research under grant ONR-OTA N00014-20-1-2205,  the Air Force Office of Scientific Research award number FA-9550-19-1-0169, and the Natural Sciences and Engineering Research Council of Canada (NSERC).

\bibliography{references}
\bibliographystyle{abbrv}

\appendix
\section{Appendix}
\subsection{BigMAC}
\BigMAC{} is a fine-grained SEPolicy static analysis tool~\cite{bigmac}. In this section we  discuss the functionality the process behind how \BigMAC{} works and what results can be generated. \BigMAC{} first walks the filesystem and extracts files' DAC file permissions, SELinux labels and Linux capabilities. Then, it parses the system's init scripts and simulates commands that affect the filesystem (e.g., \texttt{mkdir}, \texttt{chmod}, etc.) as well as \texttt{service} commands which execute service binaries. Performing boot emulation is required to create files in the \texttt{/sys}, \texttt{/dev} and \texttt{/data} directories, which would not be present in a static firmware image.

After the boot process is emulated, \BigMAC{} begins the Backing File Recovery step, where it assigns the appropriate SELinux file types and domains to all files in the extracted filesystem.
This is done by decompiling the extracted binary SEPolicy file to a multi-edge directional graph via the Access Vector rules (AVrules). Afterwards, it correlates policy types to actual files on the initialized filesystem. File objects are straightforward to correlate since their SELinux policy types are captured in the extraction step. For process subjects, Type Enforcement rules related to process transitions are inverted and processed allowing for the correlation of subject types and their executable binary backing files.

Using the full set of subject nodes, \BigMAC{} constructs a dataflow graph which simplifies the SELinux policy's access vectors into a read/write model. The dataflow graph captures all data flows allowed by AVRules for all subjects and objects by considering vectors that imply a read or a write. In the dataflow graph, objects can be one of two types: file objects and Inter-Process Communication (IPC) objects. As discussed in the previous step, file objects can contain multiple backing files, each with its own MAC/DAC/CAP metadata. On the other hand, IPC objects typically do not have any backing files and are tagged with the underlying AVClass. For instance, all classes that derive from the \texttt{socket} class are tagged as IPC objects.

The recovered subject nodes are also used in the Process Inflation step. For each subject node, \BigMAC{} attempts to match the subject node to a service definition by comparing the subject's backing file with the binary file in the service definition. If the service is enabled and is not a one-shot (transient) process, the service's defined security options are assigned to a new process. This process is then inserted into a concrete process tree.

The final step in the process is the Attack Graph Instantiation. In this step, all file objects within the dataflow graph are expanded, such that each file corresponds to one node in the graph, encompassing all of this file's MAC/DAC/CAP attributes. All of the edges to and from the original file object are duplicated for each individual node. This expanded graph is overlaid onto the concrete process tree, whereby, for each process in the process tree, all in- and out-edges in the corresponding subject in the dataflow graph are copied to the process tree. In the resultant graph, concrete processes have concrete edges to all the objects they can read from or write to. \BigMAC{} then uses this graph to generate Prolog facts that can be used for dataflow paths in that graph.


\subsection{Ground Truth Evaluation Results}
\label{appendix:ground_truth}
\begin{table}[]
\caption{Accessible sockets detected by the testing app or SAUSAGE on each of the three test devices. \ding{51} and \ding{55} indicate whether the socket was detected by the corresponding tool.}
\begin{tabular}{|lll|}
\hline
\multicolumn{1}{|l|}{\textbf{Socket}}                 & \multicolumn{1}{l|}{\textbf{Testing App}} & \textbf{SAUSAGE} \\ \hline
\multicolumn{3}{|c|}{\textbf{Motorola 7.1.1 NPIS26.48-43-2}}                                                         \\ \hline
\multicolumn{1}{|l|}{dnsproxyd}                       & \multicolumn{1}{l|}{\ding{51}}               & \ding{51}           \\ \hline
\multicolumn{1}{|l|}{fwmarkd}                         & \multicolumn{1}{l|}{\ding{51}}               & \ding{51}           \\ \hline
\multicolumn{1}{|l|}{logd}                            & \multicolumn{1}{l|}{\ding{51}}               & \ding{51}           \\ \hline
\multicolumn{1}{|l|}{logdr}                           & \multicolumn{1}{l|}{\ding{51}}               & \ding{51}           \\ \hline
\multicolumn{1}{|l|}{logdw}                           & \multicolumn{1}{l|}{\ding{51}}               & \ding{51}           \\ \hline
\multicolumn{1}{|l|}{perfd}                           & \multicolumn{1}{l|}{\ding{51}}               & \ding{51}           \\ \hline
\multicolumn{3}{|c|}{\textbf{Samsung 7.0.0 NRD90M G920FXXU5EQJ1}}                            \\ \hline
\multicolumn{1}{|l|}{dnsproxyd}                       & \multicolumn{1}{l|}{\ding{51}}               & \ding{51}           \\ \hline
\multicolumn{1}{|l|}{fwmarkd}                         & \multicolumn{1}{l|}{\ding{51}}               & \ding{51}           \\ \hline
\multicolumn{1}{|l|}{logd}                            & \multicolumn{1}{l|}{\ding{51}}               & \ding{51}           \\ \hline
\multicolumn{1}{|l|}{logdr}                           & \multicolumn{1}{l|}{\ding{51}}               & \ding{51}           \\ \hline
\multicolumn{1}{|l|}{logdw}                           & \multicolumn{1}{l|}{\ding{51}}               & \ding{51}           \\ \hline
\multicolumn{1}{|l|}{/dev/socket/jack/set.priority} & \multicolumn{1}{l|}{\ding{55}}               & \ding{51}           \\ \hline
\multicolumn{1}{|l|}{@dev/socket/jack/set.priority}   & \multicolumn{1}{l|}{\ding{55}}               & \ding{51}           \\ \hline
\multicolumn{3}{|c|}{\textbf{Samsung 8.0.0 R16NW G930UUES4CRH2}}                                                     \\ \hline
\multicolumn{1}{|l|}{dnsproxyd}                       & \multicolumn{1}{l|}{\ding{51}}               & \ding{51}           \\ \hline
\multicolumn{1}{|l|}{napproxyd}                       & \multicolumn{1}{l|}{\ding{51}}               & \ding{51}           \\ \hline
\multicolumn{1}{|l|}{fwmarkd}                         & \multicolumn{1}{l|}{\ding{51}}               & \ding{51}           \\ \hline
\multicolumn{1}{|l|}{logd}                            & \multicolumn{1}{l|}{\ding{51}}               & \ding{51}           \\ \hline
\multicolumn{1}{|l|}{logdr}                           & \multicolumn{1}{l|}{\ding{51}}               & \ding{51}           \\ \hline
\multicolumn{1}{|l|}{logdw}                           & \multicolumn{1}{l|}{\ding{51}}               & \ding{51}           \\ \hline
\multicolumn{1}{|l|}{pdx/system/vr/display/client}    & \multicolumn{1}{l|}{\ding{55}}               & \ding{51}           \\ \hline
\multicolumn{1}{|l|}{pdx/system/vr/display/manager}   & \multicolumn{1}{l|}{\ding{55}}               & \ding{51}           \\ \hline
\multicolumn{1}{|l|}{pdx/system/vr/display/vsync}     & \multicolumn{1}{l|}{\ding{55}}               & \ding{51}           \\ \hline
\multicolumn{1}{|l|}{/dev/socket/jack/set.priority}   & \multicolumn{1}{l|}{\ding{55}}               & \ding{51}           \\ \hline
\multicolumn{1}{|l|}{@dev/socket/jack/set.priority}   & \multicolumn{1}{l|}{\ding{55}}               & \ding{51}           \\ \hline
\end{tabular}
\label{table:gte}
\end{table}


Table~\ref{table:gte} shows all accessible sockets detected by either SAUSAGE or the testing app we used in the ground truth evaluation, as well as the version information of the Android devices tested. We detect more sockets than the ground truth testing app as our static analysis approach allows the detection of inactive sockets that could be created under certain conditions or configurations. The ABSTRACT socket @dev/socket/jack/set.priority was detected twice by SAUSAGE, once as a false-positive FILESYSTEM socket and the second time as a true-positive ABSTRACT socket. This false positive result is due to the peculiar construction of the socket address itself in the \textit{apaservice} binary. The address string is first set to ``/dev/socket/jack/set.priority.'' Afterwards, the `/' character at the 0th index is replaced by a null byte. This effectively changes the address from a FILESYSTEM to an ABSTRACT socket address. However, due to the imprecision of the static analysis, both addresses are reported by SAUSAGE. 

\end{document}